\documentclass[a4paper,aps,showpacs,showkeys,twocolumn]{revtex4}
\RequirePackage[english]{babel}
\RequirePackage[latin1]{inputenc}
\RequirePackage[T1]{fontenc}
\RequirePackage{mathrsfs}
\RequirePackage{amsmath}
\RequirePackage{amssymb}
\RequirePackage{amsbsy}
\RequirePackage{color}
\RequirePackage{bm}
\RequirePackage{slashed}
\begin{document}
\title{\bf{A generally-relativistic gauge classification of the Dirac fields}}
\author{Luca Fabbri}
\affiliation{INFN \& Dipartimento di Fisica, Universit{\`a} di Bologna,\\
Via Irnerio 46, 40126 Bologna, ITALY}
\date{\today}
\begin{abstract}
We consider generally-relativistic gauge transformations for the spinorial fields finding two mutually exclusive but together exhaustive classes in which fermions are placed adding supplementary information to the results obtained by Lounesto, and identifying quantities analogous to the momentum vector and the Pauli-Lubanski axial vector we discuss how our results are similar to those obtained by Wigner; by taking into account the system of Dirac field equations we will investigate the consequences for the dynamics: and in particular we shall address the problem of getting the non-relativistic approximation in a consistent way. We are going to comment on extensions.
\end{abstract}
\maketitle
\section{INTRODUCTION}
The torsional completion of gravity in the presence of electrodynamics is the most general theory in which we can exhaustively couple both spin and energy beside the current of a general spinor field; even if in practical situations it might be possible to straighten torsion and flatten curvature while setting electrodynamic fields to zero nevertheless one could still have non-trivial contributions in the connection: in fact non-inertial frames as well as Aharonov-Bohm type of effects may still be involved.

These contributions have room to absorb degrees of freedom from the spinor fields without losing generality having them restricted to specific forms: in terms of these special forms we will obtain additional information in terms of Lounesto classifications \cite{L,Cavalcanti:2014wia}; moreover, we will be able to identify quantities that can be recognized to have the same essence of the momentum vector and the Pauli-Lubanski axial vector used in Wigner classification.
\section{GENERALLY-RELATIVISTIC AND GAUGE TRANSFORMATIONS FOR DIRAC FIELDS}
We refer to \cite{A-L, L-M, Fabbri:2006xq, Fabbri:2009se, K, Fabbri:2011kq, Fabbri:2013gza, Fabbri:2014dxa} for exhaustiveness reporting only the essential assumptions starting from the fact that we consider $(1\!+\!3)$-dimensional space-times with $\frac{1}{2}$-spin spinor fields verifying the system of Dirac field equations.
\subsection{Geometrical Kinematics}
From a kinematic point of view, fields are defined as what transforms under certain transformation laws.
\subsubsection{Spinorial Transformations}
General coordinate transformations, or passive transformations, define general tensors, among which the metric tensor $g_{\alpha\rho}$ is found, and given $g_{\alpha\rho}$ and $g^{\alpha\rho}$ we can raise lower or lower upper indices; with ortho-normalization procedures it is always possible to introduce a basis of fields $\xi^{\alpha}_{a}$ for which $g_{\alpha\rho}\xi^{\alpha}_{a}\xi^{\rho}_{b}\!=\!\eta_{ab}$ where $\eta^{ab}$ is the Minkowskian matrix, because any two bases are linked by the law $\xi'^{\sigma}_{a}\!=\! \Lambda^{b}_{a}\xi^{\sigma}_{b}$ we preserve the structure of the Minkowskian matrix if $\eta\!=\!\Lambda\eta\Lambda^{T}$ and thus transformations $\Lambda$ are called Lorentz transformations in real representation, or active transformations, and $\xi_{p}^{\rho}$ and $\xi^{p}_{\rho}$ are used to pass from coordinate indices to Lorentz indices while $\eta_{ap}$ and $\eta^{ap}$ are used to raise lower and lower upper indices in Lorentz form. Lorentz transformations in real representation can be written explicitly in terms of the real generators that are given by $\sigma_{ab}\!=\!-\sigma_{ba}$ and such that $[\sigma_{ab},\sigma_{cd}]\!=\!\eta_{ad}\sigma_{bc}\!-\!\eta_{ac}\sigma_{bd}\!+\!\eta_{bc}\sigma_{ad}\!-\!\eta_{bd}\sigma_{ac}$ and with parameters $\theta_{ab}\!=\!-\theta_{ba}$ in order to obtain the expression given by $\Lambda\!=\!\exp{(\frac{1}{2}\sigma^{ab}\theta_{ab})}$ in general: in particular, it is possible to see that this expression is valid also for the complex representation. Eventually we will consider the unitary phase shift $e^{iq\alpha}$ where the label $q$ is called charge, it is also known as gauge transformation, and the complex conjugation flips the charge. Lorentz transformations in complex representation are obtained by introducing the set of Clifford matrices $\boldsymbol{\gamma}^{a}$ such that they verify
\begin{eqnarray}
&\{\boldsymbol{\gamma}^{a}\!,\!\boldsymbol{\gamma}^{b}\}\!=\! 2\eta^{ab}\boldsymbol{\mathbb{I}}
\end{eqnarray}
from which
\begin{eqnarray}
&\frac{1}{4}\!\left[\boldsymbol{\gamma}^{a}
\!,\!\boldsymbol{\gamma}^{b}\right]\!=\!\boldsymbol{\sigma}^{ab}
\end{eqnarray}
implicitly defining the matrix $\boldsymbol{\pi}$ as
\begin{eqnarray}
&\boldsymbol{\sigma}_{ab}
=-\frac{i}{2}\varepsilon_{abcd}\boldsymbol{\pi}\boldsymbol{\sigma}^{cd}
\end{eqnarray}
and with which we have the commutation relations
\begin{eqnarray}
&\{\boldsymbol{\pi},\boldsymbol{\gamma}_{a}\}=0\\
&\{\boldsymbol{\gamma}_{i},\boldsymbol{\sigma}_{jk}\}
=i\varepsilon_{ijkq}\boldsymbol{\pi}\boldsymbol{\gamma}^{q}
\end{eqnarray}
and also
\begin{eqnarray}
&[\boldsymbol{\pi},\boldsymbol{\sigma}_{ab}]=0\\
&[\boldsymbol{\gamma}_{a},\boldsymbol{\sigma}_{bc}]
=\eta_{ab}\boldsymbol{\gamma}_{c}\!-\!\eta_{ac}\boldsymbol{\gamma}_{b}\\
&[\boldsymbol{\sigma}_{ab},\boldsymbol{\sigma}_{cd}]
=\eta_{ad}\boldsymbol{\sigma}_{bc}\!-\!\eta_{ac}\boldsymbol{\sigma}_{bd}
\!+\!\eta_{bc}\boldsymbol{\sigma}_{ad}\!-\!\eta_{bd}\boldsymbol{\sigma}_{ac}
\end{eqnarray}
and
\begin{eqnarray}
&\boldsymbol{\gamma}_{a}\boldsymbol{\gamma}_{b}
\!=\!\eta_{ab} \boldsymbol{\mathbb{I}}\!+\!2\boldsymbol{\sigma}_{ab}\\
&\boldsymbol{\gamma}_{i}\boldsymbol{\gamma}_{j}\boldsymbol{\gamma}_{k}
=\boldsymbol{\gamma}_{i}\eta_{jk}-\boldsymbol{\gamma}_{j}\eta_{ik}+\boldsymbol{\gamma}_{k}\eta_{ij}
+i\varepsilon_{ijkq}\boldsymbol{\pi}\boldsymbol{\gamma}^{q}
\end{eqnarray}
telling that $\boldsymbol{\sigma}_{ab}$ are complex generators which, considered together with the same parameters $\theta_{ab}\!=\!-\theta_{ba}$ above, yield the expression given according to the following form
\begin{eqnarray}
&\boldsymbol{\Lambda}\!=\!e^{\frac{1}{2}\boldsymbol{\sigma}^{ab}\theta_{ab}}
\end{eqnarray}
as the complex representation of the Lorentz transformations; we also have the unitary phase shift that is to be considered: together they give rise to expression
\begin{eqnarray}
&\boldsymbol{S}\!=\!e^{(\frac{1}{2}\boldsymbol{\sigma}^{ab}\theta_{ab}
+iq\alpha\boldsymbol{\mathbb{I}})}
\end{eqnarray}
as spinorial transformation, and with $\boldsymbol{\gamma}_{0}$ the procedure
\begin{eqnarray}
&\overline{\psi}\!=\!\psi^{\dagger}\boldsymbol{\gamma}_{0}\ \ \ \ 
\ \ \ \ \boldsymbol{\gamma}_{0}\overline{\psi}^{\dagger}\!=\!\psi
\end{eqnarray}
is the conjugation of spinor fields, so that with the spinor and its conjugate spinor we can construct the $16$ linearly-independent bi-linear spinorial quantities
\begin{eqnarray}
&2i\overline{\psi}\boldsymbol{\sigma}^{ab}\psi\!=\!S^{ab}\\
&\overline{\psi}\boldsymbol{\gamma}^{a}\boldsymbol{\pi}\psi\!=\!V^{a}\\
&\overline{\psi}\boldsymbol{\gamma}^{a}\psi\!=\!U^{a}\\
&i\overline{\psi}\boldsymbol{\pi}\psi\!=\!\Theta\\
&\overline{\psi}\psi\!=\!\Phi
\end{eqnarray}
and we have the validity of
\begin{eqnarray}
&\psi\overline{\psi}\!\equiv\!\frac{1}{4}\Phi\boldsymbol{\mathbb{I}}
\!+\!\frac{1}{4}U_{a}\boldsymbol{\gamma}^{a}
\!+\!\frac{i}{4}S_{ab}\boldsymbol{\sigma}^{ab}
\!-\!\frac{1}{4}V_{a}\boldsymbol{\gamma}^{a}\boldsymbol{\pi}
\!-\!\frac{i}{4}\Theta \boldsymbol{\pi}
\end{eqnarray}
from which we get the relationships
\begin{eqnarray}
&\left(U_{a}\boldsymbol{\gamma}^{a}
\!+\!V_{a}\boldsymbol{\gamma}^{a}\boldsymbol{\pi}\right)\psi\equiv0\\
&\overline{\psi}\left(U_{a}\boldsymbol{\gamma}^{a}
\!+\!V_{a}\boldsymbol{\gamma}^{a}\boldsymbol{\pi}\right)\equiv0
\end{eqnarray}
as well as the relationships
\begin{eqnarray}
&S_{ab}\Phi\!+\!\frac{1}{2}\varepsilon_{abik}S^{ik}\Theta\!=\!U^{j}V^{k}\varepsilon_{jkab}\\
&S_{ab}\Theta\!-\!\frac{1}{2}\varepsilon_{abik}S^{ik}\Phi\!=\!U_{[a}V_{b]}
\end{eqnarray}
together with
\begin{eqnarray}
&S_{ik}U^{i}=\Theta V_{k}\label{P1}\\
&-\frac{1}{2}\varepsilon_{abik}S^{ab}U^{i}\!=\!\Phi V_{k}\label{L1}\\
&S_{ik}V^{i}=\Theta U_{k}\label{P2}\\
&-\frac{1}{2}\varepsilon_{abik}S^{ab}V^{i}\!=\!\Phi U_{k}\label{L2}
\end{eqnarray}
and
\begin{eqnarray}
&\frac{1}{2}S_{ab}S^{ab}\!=\!\Phi^{2}\!-\!\Theta^{2}\label{norm2}\\
&U_{a}U^{a}\!=\!-V_{a}V^{a}\!=\!\Theta^{2}\!+\!\Phi^{2}\label{norm1}\\
&\frac{1}{4}S_{ab}S_{ij}\varepsilon^{abij}\!=\!2\Theta\Phi\label{orthogonal2}\\
&V_{a}U^{a}\!=\!0\label{orthogonal1}
\end{eqnarray}
called Fierz re-arrangements and being spinor identities whose importance will be seen in the following.

In what we intend to do we consider connections whose symmetric part is uniquely defined, so that the torsion has to be completely antisymmetric and therefore dualized in terms of the Levi-Civita pseudo-tensor as
\begin{eqnarray}
&\frac{1}{6}W^{\mu}\varepsilon_{\mu\alpha\sigma\nu}\!=\!Q_{\alpha\sigma\nu}
\end{eqnarray}
in terms of a pseudo-vector $W^{\mu}$ so that 
\begin{eqnarray}
&\Gamma^{\rho}_{\alpha\beta}\!=\!\Lambda^{\rho}_{\alpha\beta}
\!+\!\frac{1}{2}Q^{\rho}_{\alpha\beta}
\label{connection}
\end{eqnarray}
with the Levi-Civita connection $\Lambda^{\rho}_{\alpha\beta}$ is the decomposition of the most general connection, while the spin-connection
\begin{eqnarray}
\Omega^{a}_{b\mu}=\xi^{\nu}_{b}\xi^{a}_{\rho}
\left(\Gamma^{\rho}_{\nu\mu}-\xi^{\rho}_{k}\partial_{\mu}\xi_{\nu}^{k}\right)\label{spinconnection}
\end{eqnarray}
is written in a form that shows how connection and spin connection are equivalent; gauge potentials $A_{\mu}$ are also introduced: finally for the spinorial connection we have
\begin{eqnarray}
&\boldsymbol{\Omega}_{\mu}
=\frac{1}{2}\Omega^{ab}_{\phantom{ab}\mu}\boldsymbol{\sigma}_{ab}
\!+\!iqA_{\mu}\boldsymbol{\mathbb{I}}\label{spinorialconnection}
\end{eqnarray}
in terms of the generator-valued spin connection and in terms of the gauge potential, both combined to exhaust the most general form of the spinorial connection.

So to summarize what has been done thus far, we may say that aside from the most general coordinate transformation, the Lorentz transformation has been written in complex representation and together with the unitary phase shift, they have been combined to form the most general spinorial transformation; because the parameters of the Lorentz transformation in real or complex forms are the same, these two transformations are simply the action on Lorentz tensors or spinor fields of the same active transformation, with the difference that for the latter also unitary phase shifts must be taken: when they are taken, the result is that the complex Lorentz transformation is completed by the unitary phase shift in such a way that they eventually give rise to the most general spinorial transformation. This is best seen in the connections, because the spin connection can be complex-valued but it is only when also the gauge connection is added that together they saturate the most general spinorial connection; this is intriguing, because it suggests that the gauge connections has nothing less than the spin connection in terms of geometric necessity. Some may wish to regard this as a sort of conceptual geometric unification.
\subsubsection{Classifications}
So far we have established the general structure of the kinematical tools we will employ for the classification and next we will actually perform such a classification.

We begin by considering the special case of spinor given when $\overline{\psi}\psi \!=\!i\overline{\psi}\boldsymbol{\pi}\psi\!=\!0$ hold; for this $2i\overline{\psi}\boldsymbol{\sigma}_{ab}\psi$ can be written keeping separated its time-space and space-space components with (\ref{norm2}, \ref{orthogonal2}) then telling that these components are two $3$-dimensional vectors orthogonal with equal norm; hence it is always possible to perform three independent rotations bringing these two vectors aligned with two assigned axes, as for instance the space-space and time-space components respectively aligned with the first and second axis: therefore the most general spinor that is to respect these constraints is given by either
\begin{eqnarray}
&\!\!\psi\!=\!\phi e^{i\xi}\!\left(\!\begin{tabular}{c}
$\cos{\frac{\theta}{2}}$\\
$0$\\
$0$\\
$\pm\sin{\frac{\theta}{2}}$
\end{tabular}\!\right)\ \ \mathrm{or}\ \ \psi\!=\!\phi e^{i\xi}\!\left(\!\begin{tabular}{c}
$0$\\
$\cos{\frac{\theta}{2}}$\\
$\pm\sin{\frac{\theta}{2}}$\\
$0$
\end{tabular}\!\right)
\end{eqnarray}
although being the third axis reflection of each other, it will be enough to consider only one; we no longer have the possibility to perform boosts along the first and second axis, but we may consider the boost along the third axis with rapidity $\varphi$ given according to the form
\begin{eqnarray}
&\boldsymbol{S}_{B3}\!=\!\left(\begin{array}{cccc}
\!e^{-\frac{\varphi}{2}}\!&\!0\!&\!0\!&\!0\!\\ 
\!0\!&\!e^{\frac{\varphi}{2}}\!&\!0\!&\!0\!\\ 
\!0\!&\!0\!&\!e^{\frac{\varphi}{2}}\!&\!0\!\\ 
\!0\!&\!0\!&\!0\!&\!e^{-\frac{\varphi}{2}}\!\\
\end{array}\right)
\end{eqnarray}
showing that such form for the spinor is an eigenstate of this operator with eigenvalues that are functions with opposite rapidities: this boost can be used to normalize the above spinors in order to have them in the form
\begin{eqnarray}
&\!\!\psi\!=\!e^{i\xi}\!\left(\!\begin{tabular}{c}
$\cos{\frac{\theta}{2}}$\\
$0$\\
$0$\\
$\pm\sin{\frac{\theta}{2}}$
\end{tabular}\!\right)
\end{eqnarray}
although with the transformation $\psi\!\rightarrow\!\boldsymbol{\pi}\psi$ the two forms can be obtained from one another, suggesting that just a single form is necessary: we also have the unitary phase shift that can be used and which can always be chosen to have the unitary phase shifted away, and so it is always possible to choose a frame where the spinor has form
\begin{eqnarray}
&\!\!\psi\!=\!e^{i\xi}\!\left(\!\begin{tabular}{c}
$\cos{\frac{\theta}{2}}$\\
$0$\\
$0$\\
$\sin{\frac{\theta}{2}}$
\end{tabular}\!\right)
\end{eqnarray}
and if the spinor were charged $\xi\!=\!0$ could be chosen.

If the above two constraints do not hold then we are in the most general of the situations; this time we take into account $\overline{\psi}\boldsymbol{\gamma}_{a}\psi$ and $\overline{\psi}\boldsymbol{\gamma}_{a}\boldsymbol{\pi}\psi$ written keeping separated their time and space components with (\ref{norm1}) telling that the vector is time-like; so we can perform three independent boosts bringing its space components to vanish, with the additional constraint for which we can always perform two independent rotations bringing the axial vector to have space components aligned with one given axis, as for instance the third axis: the most general spinor respecting these constraints is given according to either
\begin{eqnarray}
&\!\!\psi\!=\!\phi e^{i\xi}\!\left(\!\begin{tabular}{c}
$e^{i\frac{\varphi}{2}}$\\
$0$\\
$\pm e^{-i\frac{\varphi}{2}}$\\
$0$
\end{tabular}\!\right)\ \ \mathrm{or}\ \ \psi\!=\!\phi e^{i\xi}\!\left(\!\begin{tabular}{c}
$0$\\
$e^{i\frac{\varphi}{2}}$\\
$0$\\
$\pm e^{-i\frac{\varphi}{2}}$
\end{tabular}\!\right)
\end{eqnarray}
although being the third axis reflection of each other, it will be enough to consider only one; by considering that the two rotations we have already used can always be chosen as those around the first and second axis, it follows that we may consider the rotation around the third axis with angle $\theta$ given according to the form
\begin{eqnarray}
\boldsymbol{S}_{R3}\!=\!\left(\begin{array}{cccc}
\!e^{i\frac{\theta}{2}}\!&\!0\!&\!0\!&\!0\!\\ 
\!0\!&\!e^{-i\frac{\theta}{2}}\!&\!0\!&\!0\!\\ 
\!0\!&\!0\!&\!e^{i\frac{\theta}{2}}\!&\!0\!\\ 
\!0\!&\!0\!&\!0\!&\!e^{-i\frac{\theta}{2}}\!\\
\end{array}\right)
\end{eqnarray}
showing that such form for the spinor is an eigenstate of this operator with eigenvalues that are functions with opposite angles: this rotation can be used to have the unitary phase shifted away in the above spinors in order to have them written in the following form
\begin{eqnarray}
&\!\!\psi\!=\!\phi\!\left(\!\begin{tabular}{c}
$e^{i\frac{\varphi}{2}}$\\
$0$\\
$\pm e^{-i\frac{\varphi}{2}}$\\
$0$
\end{tabular}\!\right)
\end{eqnarray}
although with the transformation $\psi\!\rightarrow\!\boldsymbol{\pi}\psi$ the two forms can be obtained from one another, suggesting that just a single form is necessary: in this case no unitary phase shift can be performed since the unitary phase has already been removed, and thus it is always possible to choose a frame in which the spinor has form
\begin{eqnarray}
&\!\!\psi\!=\!\phi\!\left(\!\begin{tabular}{c}
$e^{i\frac{\varphi}{2}}$\\
$0$\\
$e^{-i\frac{\varphi}{2}}$\\
$0$
\end{tabular}\!\right)
\end{eqnarray}
in the most general situation that is possible.

Summarizing, spinors come in two mutually exclusive and together exhaustive classes: The first class is given when $\overline{\psi}\psi \!=\!i\overline{\psi}\boldsymbol{\pi}\psi\!=\!0$ and in this case it is always possible to choose a frame in which the spinor has the form
\begin{eqnarray}
&\!\!\psi\!=\!e^{i\xi}\!\left(\!\begin{tabular}{c}
$\cos{\frac{\theta}{2}}$\\
$0$\\
$0$\\
$\sin{\frac{\theta}{2}}$
\end{tabular}\!\right)\label{spinorreduced}
\end{eqnarray}
with $\xi\!=\!0$ if the spinor is charged. The second class is the most general one and in this instance it is always possible to choose a frame in which the spinor has the form
\begin{eqnarray}
&\!\!\psi\!=\!\phi\!\left(\!\begin{tabular}{c}
$e^{i\frac{\varphi}{2}}$\\
$0$\\
$e^{-i\frac{\varphi}{2}}$\\
$0$
\end{tabular}\!\right)\label{spinor}
\end{eqnarray}
which is valid regardless the charge of the spinor field.

For the spinors of type (\ref{spinorreduced}) we have that
\begin{eqnarray}
&S^{02}\!=\!\sin{\theta}\\
&S^{23}\!=\!\sin{\theta}\\
&V^{0}\!=\!-\cos{\theta}\\
&V^{3}\!=\!\cos{\theta}\\
&U^{0}\!=\!1\\
&U^{3}\!=\!-1\\
&\Theta\!=\!0\\
&\Phi\!=\!0
\end{eqnarray}
displaying relationships between the vector and the axial vector; for spinors of type (\ref{spinor}) we have
\begin{eqnarray}
&S^{03}\!=\!2\phi^{2}\sin{\varphi}\\
&S^{12}\!=\!2\phi^{2}\cos{\varphi}\\
&V^{3}\!=\!2\phi^{2}\\
&U^{0}\!=\!2\phi^{2}\\
&\Theta\!=\!2\phi^{2}\sin{\varphi}\\
&\Phi\!=\!2\phi^{2}\cos{\varphi}
\end{eqnarray}
with relationships among the spin tensor and the scalars.

It is important to remark that because spinor fields are point-dependent, then also the bi-linear spinor quantities are point-dependent, and finding frames where some of them vanished could only be possible if also those frames were point-dependent; finding spinorial transformations mapping the spinor fields into reduced forms could only be possible if also those spinorial transformations were point-dependent, and that is if also those spinorial transformations were locally defined: thus such a classification can only be possible within a generally-relativistic gauge geometric environment. This situation can be interpreted by thinking that it is in terms of transformation laws that spinors are defined and because the frame is not chosen they contain redundant information beside the independent degrees of freedom: for example, although spinors with the first and the third components are the third-axis reflection of spinors with the second and the fourth components, until the third axis is assigned in general all four components must be allowed. And here above we have exploited transformations with local structure to absorb the redundant information away from the spinor field so to leave it with the essential degrees of freedom alone.

This is what we found here in addition to the Lounesto classification: Lounesto performs a mathematical categorization of the components of the spinor while we look for a physical categorization of the degrees of freedom of the spinorial field. The Cavalcanti classification \cite{Cavalcanti:2014wia} giving the degrees of freedom of the spinorial field but without employing local transformations can be placed in between.

Finally, we notice that identity (\ref{L1}) written as
\begin{eqnarray}
&\Phi V_{k}\!=\!\frac{1}{2}\varepsilon_{kabi}S^{ab}U^{i}
\end{eqnarray}
is closely resemblant to the Pauli-Lubanski axial vector
\begin{eqnarray}
&L_{k}\!=\!\frac{1}{2}\varepsilon_{kabi}M^{ab}P^{i}
\end{eqnarray}
in terms of $M^{ab}$ and $P^{i}$ considered to be the generators of roto-translations in the Poincar\'{e} group: one important property of the Pauli-Lubanski axial vector is that it verifies 
$L_{k}P^{k}\!=\!0$ and from (\ref{orthogonal1}) it is $V_{a}U^{a}\!=\!0$ showing even more similarities between the geometrical objects used in this classification and the quantum operators employed in the Wigner classification; in the Wigner classification, Casimir operators are $P_{k}P^{k}$ and $L_{k}L^{k}$ as those of the full Poincar\'{e} group. It is our aim now to consider the Wigner classification involving quantum states and have it compared to a classification valid for spinors: we will see that despite the obvious differences, there are also analogies quite important as well. We discuss them in parallel.

Wigner classification for Poincar\'{e} Casimir operators is
\begin{eqnarray}
&P_{k}P^{k}\!=\!m^{2}>0\\
&L_{k}L^{k}\!=\!-m^{2}(s\!+\!1)s
\end{eqnarray}
showing that Casimir operators are the momentum and the spin of the particle with $m$ and $s$ being the mass and spin quantum numbers, and this is to be compared to the identities (\ref{norm1}) and (\ref{P1}, \ref{L1}) again with (\ref{norm1}) resulting into
\begin{eqnarray}
&U_{a}U^{a}\!=\!\Theta^{2}\!+\!\Phi^{2}>0\\
&(\Phi^{2}\!-\!\Theta^{2})V_{k}V^{k}\!=\!-(\Theta^{2}\!+\!\Phi^{2})\frac{1}{2}S^{ab}S_{ab}
\end{eqnarray}
in terms of bi-linear vector and axial vector together with the antisymmetric tensor fields; Wigner classification for Poincar\'{e} Casimir operators in the case of masslessness is given for $P_{k}P^{k}\!=\!L_{k}L^{k}\!=\!0$ with $L_{k}\!=\!hP_{k}$ which shows that the two Casimir operators are the momentum and the helicity of the particle and that they are proportional with $h$ being the helicity quantum number as their proportionality factor, similar to the case of the first class where $U_{k}U^{k}\!=\!V_{k}V^{k}\!=\!0$ with $V_{k}\!=\!-\cos{\theta}U_{k}$ showing that bi-linear vector and axial vector are proportional indeed.

For boost eigen-spinors (\ref{spinorreduced}) we could not vanish the third component of the bi-linear vector, suggesting that the bi-linear vector encodes information about velocity, compatibly with the fact that in the Wigner classification the role of the momentum vector is played by the bi-linear vector, while for rotation eigen-spinors (\ref{spinor}) we could not vanish the third component of the bi-linear axial vector, suggesting that the bi-linear axial vector encodes information about spin, and compatibly with the fact that in the Wigner classification the role of the Pauli-Lubanski axial vector is played by the bi-linear axial vector.

Again for the comparison of the approaches, what we have done here and the well known Wigner classification have analogies, but also an important difference in the fact that we do not make use of the concept of mass as instead Wigner does: here we have a kinematic categorization based on bi-linear spinor quantities while Wigner has a dynamical categorization based on particle momenta.

So we have Lounesto, this and Wigner classifications in order of increasing proximity to physical dynamics.

As we have said, that spinors display a peculiar form without for that being less general is due to the fact that we have not removed degrees of freedom but only their redundant information due to the frame has been transferred into the frame itself; and as we have already remarked in the previous section these frames can be local and non-inertial frames correspond to non-trivial contributions within spinorial connections: consequently such a preservation of degrees of freedom is clearest when spinorial covariant derivatives are taken into consideration.

Spinorial dynamics is what we will consider next.
\subsection{Differential Dynamics}
The spinorial dynamics is encoded by the field equations and three conserved quantities given by energy and spin, together with the current: the energy is in general a non-symmetric tensor but after torsion is decomposed and a Belinfante procedure is applied we obtain that the kinetic energy is given by the symmetric tensor
\begin{eqnarray}
&E^{\rho\sigma}\!=\!\frac{i}{4}(\overline{\psi}\boldsymbol{\gamma}^{\rho}\boldsymbol{\nabla}^{\sigma}\psi
\!-\!\boldsymbol{\nabla}^{\sigma}\overline{\psi}\boldsymbol{\gamma}^{\rho}\psi+\\
\nonumber
&+\overline{\psi}\boldsymbol{\gamma}^{\sigma}\boldsymbol{\nabla}^{\rho}\psi
\!-\!\boldsymbol{\nabla}^{\rho}\overline{\psi}\boldsymbol{\gamma}^{\sigma}\psi)\label{energy}
\end{eqnarray}
while the completely antisymmetric spin can be dualized in terms of the completely antisymmetric Levi-Civita pseudo-tensor getting the following form
\begin{eqnarray}
&S^{\rho}\!=\!X\overline{\psi}\boldsymbol{\gamma}^{\rho}\boldsymbol{\pi}\psi\label{spin}
\end{eqnarray}
in terms of an axial vector, with the electric current being given in terms of the usual expression
\begin{eqnarray}
&J^{\mu}\!=\!q\overline{\psi}\boldsymbol{\gamma}^{\mu}\psi\label{current}
\end{eqnarray}
closing the set of conserved quantities; conserved quantities alone are not enough for a complete description of the dynamical information and in order to determine the dynamical behaviour of the Dirac field we need to assign
\begin{eqnarray}
&i\boldsymbol{\gamma}^{\mu}\boldsymbol{\nabla}_{\mu}\psi
\!-\!XW_{\sigma}\boldsymbol{\gamma}^{\sigma}\boldsymbol{\pi}\psi
\!-\!m\psi\!=\!0
\end{eqnarray}
with the most general form of the torsional interactions and they are known to be called Dirac field equations.

To analyze what happens to the classes discussed above from a dynamical perspective, we start by noticing some weird circumstances related to the spinor of the first class represented by (\ref{spinorreduced}) in general: as it can be seen with a direct substitution, it has 
$U_{a}\!=\!(1,0,0,1)$ with the consequence that the velocity density is constant and therefore any integration over infinite volumes would give rise to infinite results, which look awkward in themselves but even worse is the fact that they would give rise to infinite values of conserved quantities; that in this class spinors are kinematically possible but dynamically problematic is why, despite they are interesting \cite{daRocha:2013qhu,daSilva:2012wp,Ablamowicz:2014rpa, daRocha:2016bil,daRocha:2008we,Villalobos:2015xca,Cavalcanti:2014uta}, nevertheless we will no longer consider them in what is following.

Rather we will focus on the spinors of the second class represented by expression (\ref{spinor}) in general: energy is
\begin{eqnarray}
\nonumber
&E_{\mu\nu}\!=\!\frac{1}{4}(\partial_{\mu}\varphi V_{\nu}\!+\!
\partial_{\nu}\varphi V_{\mu})+\\
&+\frac{1}{8}(\Omega^{ab}_{\phantom{ab}\mu}\varepsilon_{\nu abk}
\!+\!\Omega^{ab}_{\phantom{ab}\nu}\varepsilon_{\mu abk})V^{k}
\!-\!\frac{q}{2}(A_{\mu}U_{\nu}\!+\!A_{\nu}U_{\mu})
\end{eqnarray}
with the spin given by
\begin{eqnarray}
&S^{\rho}\!=\!XV^{\rho}
\end{eqnarray}
and the electric current as
\begin{eqnarray}
&J^{\mu}\!=\!qU^{\mu}
\end{eqnarray}
as the conserved quantities; the field equations
\begin{eqnarray}
\nonumber
&\left[(\frac{1}{2}\partial_{\mu}\varphi
\!+\!\frac{1}{4}\Omega^{\alpha\nu\rho}\varepsilon_{\alpha\nu\rho\mu}
\!-\!XW_{\mu})\boldsymbol{\gamma}^{\mu}\boldsymbol{\pi}+\right.\\
&\left.+(i\partial_{\mu}\ln{\phi}\!+\!\frac{i}{2}\Omega^{a}_{\phantom{a}\mu a}
\!-\!qA_{\mu}) \boldsymbol{\gamma}^{\mu}\!-\!m\mathbb{I}\right]\psi\!=\!0
\end{eqnarray}
constitute the explicit form of the Dirac field equations.

We must notice that all conserved quantities are given in terms of the vector and axial vector, and in particular the electric current and the axial spin are exactly the vector and axial vector up to a multiplicative factor given by the coupling constants; the vector and the axial vector respectively describe the velocity vector and the spin axial vector, and they are respectively analogous to the momentum vector and the Pauli-Lubanski axial vector in Wigner classification: as such they have to be regarded as the fundamental quantities not only in classifying fields but also in the determination of the field dynamics.

Also notice that the Dirac field equations in absence of torsion-gravity and electrodynamics are still in the form
\begin{eqnarray}
\nonumber
&\left[\frac{1}{2}(\partial_{\mu}\varphi
\!+\!\frac{1}{2}\Omega^{\alpha\nu\rho}\varepsilon_{\alpha\nu\rho\mu})
\boldsymbol{\gamma}^{\mu}\boldsymbol{\pi}+\right.\\
&\left.+i(\partial_{\mu}\ln{\phi}\!+\!\frac{1}{2}\Omega^{a}_{\phantom{a}\mu a}) \boldsymbol{\gamma}^{\mu}\!-\!m\mathbb{I}\right]\psi\!=\!0
\end{eqnarray}
and despite that $\Omega^{a}_{\phantom{a}b\mu}$ may have zero curvature and thus they may be integrable nevertheless they may still be non-vanishing; we also said that spinorial connections with non-trivial contributions contain information about frames that are non-inertial, and these non-inertial effects are the result of transferring redundant information away from the spinor fields: now we can see in the clearest way that information can be transferred between local spinor fields and non-trivial spinorial connections while leaving both terms within the parentheses completely invariant.

As an explicit example consider plane-wave solutions normalized as to be $\overline{\psi}\psi\!=\! 2\!\neq\!0$ so that spinors belong to the second class; the above analysis can be performed, and after boosting into the rest frame and aligning the momentum along the third axis we have that
\begin{eqnarray}
\psi\!=\!e^{-imt}\!\left(\begin{array}{c}
1\\
0\\
1\\
0
\end{array}\right)\label{pws}
\end{eqnarray}
as the most we can reduce if only global Lorentz and spinor transformations are taken; if we were to allow also local Lorentz and spinor transformations then we could also perform a rotation of angle $\theta\!=\!2mt$ getting
\begin{eqnarray}
\psi\!=\!\left(\begin{array}{c}
1\\
0\\
1\\
0
\end{array}\right)\label{pwsr}
\end{eqnarray}
being (\ref{spinor}) when $\phi\!=\!1$ and $\varphi\!=\!0$ hold. Nevertheless, with global spinorial transformations the spinorial connection would be zero; in the case of local spinorial transformations the spinorial connection (\ref{spinorialconnection}) would be given in the form $\boldsymbol{\Omega}_{\mu}\!=\!2m\partial_{\mu}t\boldsymbol{\sigma}_{12}$ implying $\Omega^{12}_{\phantom{12}t}\!=\!2m$ and so that the spinorial connection would contain information about the non-inertial frames. The global transformations can have redundant information re-arranged into the frame, but local components can be re-arranged only into non-inertial frames corresponding to non-trivial contributions within spinorial connections, and there will be preservation of degrees of freedom within spinorial covariant derivatives.

To conclude we notice that in this example plane-wave solutions gave rise to constant spinors, and now we want to prove that such circumstance is general: by imposing the condition $i\boldsymbol{\nabla}_{\mu}\psi\!=\!P_{\mu}\psi$ for the spinorial covariant derivatives of the spinor (\ref{spinor}) we get the relationships
\begin{eqnarray}
&\left[\frac{i}{2}\Omega^{ab}_{\phantom{ab}\mu}\boldsymbol{\sigma}_{ab}
\!+\!\frac{1}{2}\partial_{\mu}\varphi\boldsymbol{\pi}
\!+\!(i\partial_{\mu}\ln{\phi}\!-\!P_{\mu})\mathbb{I}\right]\!\psi\!=\!0
\end{eqnarray}
from which $\partial_{\mu}\varphi\!=\!\partial_{\mu}\phi\!=\!0$ are obtained. What this would imply is that for any momentum the spinor is constant so that its bi-linear spinorial quantities are constant densities, and their integrals give infinite conserved quantities.

In the following we are going to give general comments about all of these results and some interpretation.
\section{NON-RELATIVISTIC CORRESPONDENCE}
So far we studied how generally-relativistic and gauge transformations could be used to have spinors reduced into forms classifiable in two classes, even if a subsequent investigation made it clearer that those belonging to the first class were singular and only those of the second class received further attention; we have highlighted the fact that such a classification was based on the utilization of the velocity vector and spin axial vector, analogous to the momentum vector and the Pauli-Lubanski axial vector of Wigner classification, and being the conserved quantities with which the dynamics is described. As it seems, the velocity vector $U^{\mu}$ and spin axial vector $V^{\mu}$ appear to be fundamental entities in the description of spinorial fields.

In the following we will consider what happens in the case of the non-relativistic approximation, highlighting the role played by the spin tensor in such limit.
\subsection{Spin}
Here above, spinor fields were defined in terms of spinorial transformations and initially they had $8$ independent components; because spinorial transformations are a representation of the Lorentz group with $6$ parameters, we had spinor fields reduced to $2$ independent components.

In the non-relativistic case, we have no possibility to define the boosts: semi-spinor fields would be defined in terms of semi-spinorial transformations and initially they would have $4$ independent components; because semi-spinorial transformations are a representation of the rotation group with $3$ parameters, we would have semi-spinor fields reduced to $1$ independent component alone.

What this means is that in the non-relativistic approximation, starting from spinors to get semi-spinors the extra degree of freedom must be lost somehow.

With form (\ref{spinor}) in standard representation we get
\begin{eqnarray}
&\frac{\phi}{\sqrt{2}}\!\left(\!\begin{tabular}{c}
$e^{i\frac{\varphi}{2}}\!+\!e^{-i\frac{\varphi}{2}}$\\
$0$\\
$e^{i\frac{\varphi}{2}}\!-\!e^{-i\frac{\varphi}{2}}$\\
$0$
\end{tabular}\!\right)\!=\!\sqrt{2}\phi\!\left(\!\begin{tabular}{c}
$\cos{\frac{\varphi}{2}}$\\
$0$\\
$i\sin{\frac{\varphi}{2}}$\\
$0$\\
\end{tabular}\!\right)
\end{eqnarray}
so that we can see that the non-relativistic semi-spinor 
\begin{eqnarray}
&\phi\!\left(\!\begin{tabular}{c}
$1$\\
$0$\\
\end{tabular}\!\right)
\end{eqnarray}
coincides with either the upper or the lower part of the above spinor only if $\sin{\varphi}\!=\!0$ which is to be regarded as the constraint that makes the degrees of freedom decrease from two to one alone: in this case we obtain that conditions $\Theta\!=\!0$ and $\Phi\!\neq\!0$ hold. Thus for the moment we interpret these two conditions together with the vanishing of the spatial part of the velocity density vector to be what encodes the non-relativistic approximation.

For spinors of the second class, albeit possible to find a frame in which the velocity loses the entire spatial part, this does not mean that the momentum will also lose the whole space component; this situation is easily understood by thinking that internal structures are present, due to the presence of spin: in fact by taking into account the momentum $i\boldsymbol{\nabla}_{\alpha}\psi\!=\!P_{\alpha}\psi$ the Gordon decomposition
\begin{eqnarray}
&P^{\alpha}\Phi\!=\!mU^{\alpha}\!+\!\frac{1}{2}(\nabla_{\mu}S^{\mu\alpha}
\!+\!XW_{\sigma}S_{\mu\nu}\varepsilon^{\sigma\mu\nu\alpha})
\end{eqnarray}
shows that velocity and momentum are not even aligned if the spin tensor is present. As it is widely known and according to what we have demonstrated before plane-wave solutions give rise to bi-linear spinor densities that are constant, yielding integral quantities that are infinite and so meaningless; but even if we were to allow plane-wave solutions in some localized spatial region so to have a spin tensor constant yielding conserved quantities finite and thus meaningful, nevertheless only when torsion is vanishing $P^{\alpha}\Phi\!=\!mU^{\alpha}$ holds. What this shows us is that the non-relativistic limit is a regime in which beside small velocities also some information about spin is required.

As a matter of fact, for (\ref{spinor}) it is possible to see that conditions $\Theta\!=\!0$ and $\Phi\!\neq\!0$ are equivalent to assuming that the spin tensor $S_{ij}$ loses its space-time components without losing the space-space components: because this condition requires the splitting in spatial and temporal parts it is not an invariant condition, but at least it serves to show that the meaning of the non-relativistic limit is that of vanishing the part of the spin tensor related to boosts without vanishing the part of the spin tensor related to rotations. The non-relativistic limit is a regime requiring in addition to small spatial part of the velocity vector also small space-time part of the spin tensor.

It is to be noticed however that the spin tensor is not the spin axial vector, and although they are related nevertheless only the spin axial vector is the conserved quantity that appears in the field equations.
\section{COMPLETIONS}
Up until now, we have worked in a generally-relativistic and gauge covariant environment categorizing spinors in two classes by employing the velocity vector and spin axial vector; these two vectors have also been seen to give rise to all conserved quantities with which the dynamics is described eventually. And finally, we have also discussed the criteria for the consistent non-relativistic limit.

The spin is the most essential quantity throughout the entire discussion: it is the spin axial vector what allows to complete the set of tools needed to classify spinors and the system of conserved quantities needed to determine the dynamics; and it is the spin tensor what renders non-trivial the limit in non-relativistic approximation.

In generally relativistic theories the spin axial vector is coupled to torsion, and thus one would expect a similarly fundamental role of the torsion tensor itself.

Discussions about the torsion coupling to spin and its effects have been studied in \cite{Fabbri:2011mg, Fabbri:2014vda, Fabbri:2014iya, Fabbri:2015xga}.
\section{CONCLUSION}
In this paper, we have been discussing how local active Lorentz and gauge transformations can be used in order to define spinors and to eventually reduce them to the very specific forms (\ref{spinorreduced}, \ref{spinor}) remarking that this gives supplementary information to the Cavalcanti and Lounesto classifications, and we noticed analogies and differences with Wigner classification: Lounesto did a mathematical analysis while ours is a physical discussion based on kinematical quantities and Wigner's is a physical discussion based on dynamical quantities. In studying the dynamics, we have highlighted that the first class might be problematic and focusing on the second class, we have seen in what way the velocity vector and the spin axial vector were the objects in terms of which all conserved quantities could be written, increasing their importance as fundamental entities in the description of spinorial fields.

We also discussed that in the non-relativistic regime small velocities alone are not enough and also constraints on the spin tensor are needed: we have been seeing that the non-relativistic limit is given for small spatial part of the velocity vector and space-time part of the spin tensor.

As the spin is coupled to torsion, we inferred that torsion should also be an important aspect to consider.

Opportunities for more investigations are in terms of studying whether the correspondence to Wigner quantum methods is only formal or truly conceptual; and having spinors coupled to torsion or with the presence of non-trivial topological terms \cite{daRocha:2007sd,Rodrigues:2005yz,daRocha:2011yr} can foster their dynamics with properties that may have considerable influence.

Studying these dynamical properties is yet another opportunity for further investigations on this subject.

We will leave this task to some following work.
\begin{acknowledgments}
I would like to thank very warmly Professor Rold\~{a}o da Rocha for the interesting discussions we had had about the subject of spinor field classification during the SUSY 2013 meeting, held at SISSA, in Trieste, Italy.

Also I would like to thank the referee for directing my attention toward important references.
\end{acknowledgments}

\end{document}